# From the ideal gas to an ideal glass: a thermodynamic route to random close packing.

**Leslie V. Woodcock**

Department of Chemical and Biomolecular Engineering
National University of Singapore, Singapore 117576
e-mail chewlv@nus.edu.sg

A random close-packed (RCP) state of spheres, was first investigated by Bernal [1] as a model of simple liquids. The hard-sphere model plays a central role in condensed matter physics; accordingly its amorphous solid state RCP has since been well-characterized in granular laboratories [2-5], by mathematical analysis [6], and computer experiments [7-9]. Yet, fundamental questions like "What is random packing?" originally asked by Nature's correspondent [10], are still being asked in the scientific literature; questions such as "Is random close packing well-defined?" [11] or "Why is random close packing reproducible?" [12]. Here, we first obtain a reproducible RCP state by a well-defined rate process, which is first-order in free volume, starting from an equilibrium thermodynamic state of the hard-sphere fluid. We then report the RCP state is also reproduced by a thermodynamic pathway that inhibits the frequent production of small crystal nucleites [13], which can lead to heterogeneities at close packing. The simple-cubic single-occupancy (SCSO) cell system, in which crystallization is de facto prohibited, provides a reversible path all the way from the ideal gas to the "ideal glass" (RCP). By such a route, unique characterization of RCP becomes possible; the volume fraction is found to be 0.6333$\pm$0.0005 with a residual entropy $\Delta S$= 1.08$\pm$0.02 $Nk_B$ (N is number of spheres and $k_B$ is Boltzmann's constant).

Figure 1 shows a characteristic radial distribution function (RDF) of an RCP state This RDF is identical that of RCP states produced by irreversible compaction processes, but it has been obtained here from a reversible pathway to RCP that utilizes a single occupancy constraint of sphere centers to a local cell.

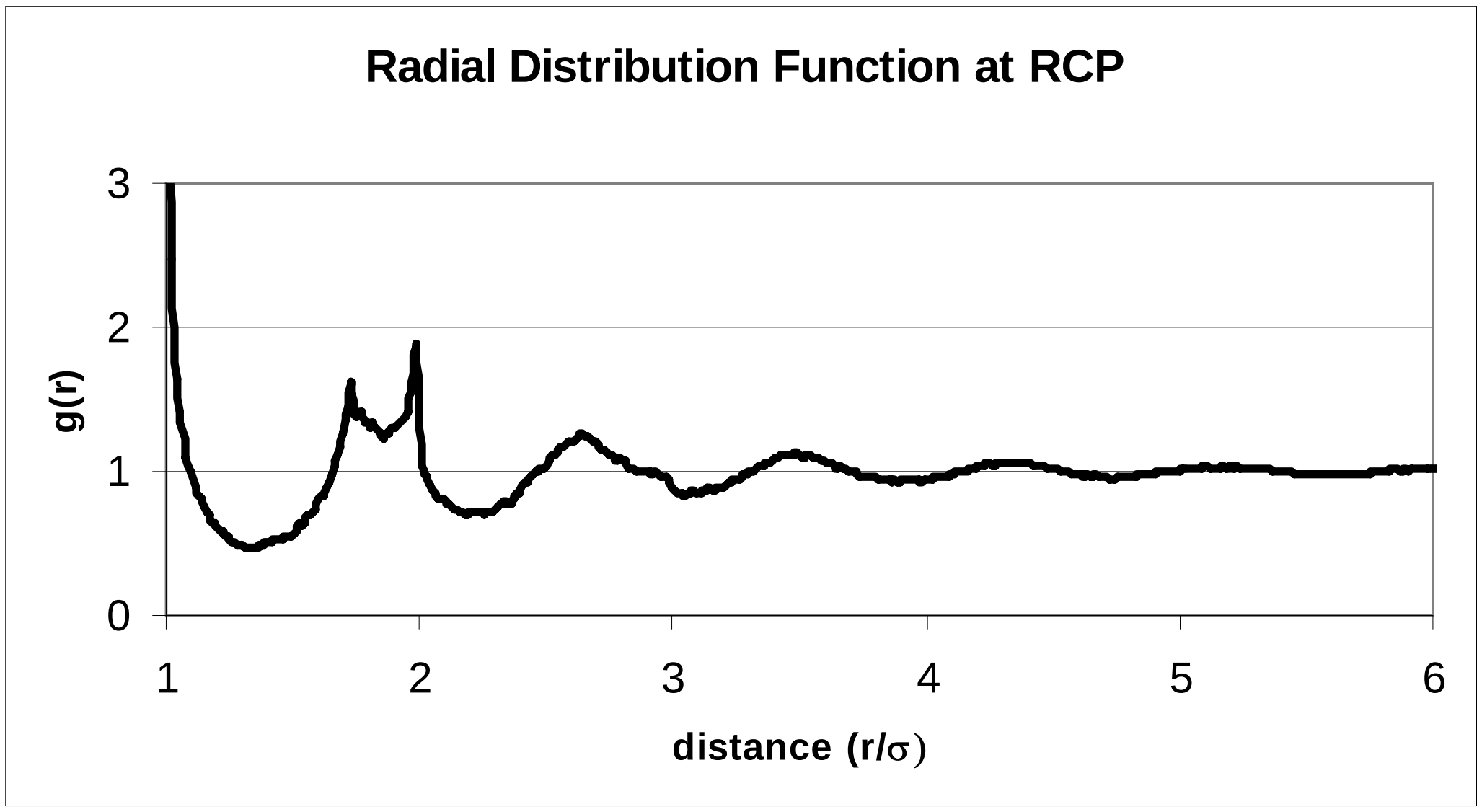

**Figure 1 Radial distribution function (RDF) of the RCP state at the density $\rho\sigma^3$=1.2155. The structure is characterized by two spiked peaks at $r/\sigma = 3^{1/2}$ and 2, by an absence of a peak at crystal distance $r/\sigma = 2^{1/2}$, and by randomness (g(r) =1) in pair correlations beyond ~ 5 σ.**

Structural properties such as distribution functions [8], or order parameters [11] may be used to characterize the state, but equilibrium states of matter are defined only by fixing macroscopic state and process variables. For single homogeneous phases, usually temperature and pressure define the state, the thermodynamic properties such as enthalpy or entropy, are single-valued state functions. Most composite solid materials, and all glasses, exhibit thermodynamic properties which are not single-valued state functions. Metastable solid states are only as well-defined as both the process the produces them, and the initial equilibrium state from which it began.

For spheres, one could define either continuous uniaxial compaction processes, analogous to the real mesoscale processes of sedimentation, filtration or slip casting, or, an idealized batch process isothermal isotropic compaction, as in the thermal quenching of atomic or molecular liquids to glasses.

Either way, any RCP state must be defined by :-

(i) the temperature (or pressure) at which it exists; i.e. absolute zero, or equally, for hard spheres, infinite pressure, i.e. $k_BT/p\sigma^3 = 0$ ; ($k_B$ is Boltzmann's constant)

(ii) one or more rate constants that completely define the compaction process,

(iii) the initial state from which compaction began; this must be an equilibrium fluid state, i.e. at a density less than the fluid freezing density of $\rho_f=0.95\sigma^3$ ,

(iv) and N, the number of spheres; all small finite systems exhibit thermodynamic properties with N-dependence that varies with type of boundary conditions, but which converge to a limiting value for $N \rightarrow \infty$ .

Compaction processes zeroth-order in free volume (or pressure) , i.e. $d\sigma/dt$ is held constant, miss the "RCP window" . If the rate constant is high enough to avoid nucleation, it is too high for re-equilibration onto the RCP state [11] . Here we define an isothermal compaction process by a rate law which is first-order in free volume.

$$dV/dt = - k_1 (V(t) - V_\infty) \qquad \text{equation (1)}$$

V is the volume of a cube containing N spheres with periodic boundary conditions. $V_\infty$ is the limiting close-packed state at $kT/p\sigma^3 = 0$, and is *a priori* unknown, but can be predicted in any time $\Delta t$ from the mean pressure using the self-consistent free volume (SCFV) equation-of-state

$$V_\infty = \langle V(\Delta t)\rangle \, [\, 1-1/\langle Z(\Delta t)\rangle]^3 \qquad \text{equation (2)}$$

where $Z = pV/ Nk_BT$ and the angular brackets denote an average over $\Delta t$.

If V is fixed, the sphere expansion rate $d\sigma/dt$, at any time t, is

$$d\sigma/dt = k_1 \sigma (1- V_\infty /V(t)) / 3 \qquad \text{equation (3)}$$

and is incorporated continuously into the equations-of-motion. The NEMD simulation obtains the collision sequence by solving the quadratic equation for the time (t) to the next collision of any pair of spheres i and j

$$0 = r_{ij}^2 - \sigma^2 + 2t ( \underline{\mathbf{r}}_{ij}\underline{\mathbf{v}}_{ij} - d\sigma/dt) + t^2 [ \underline{\mathbf{v}}_{ij}^2 -(d\sigma/dt )^2 ] \qquad \text{equation (4)}$$

Upon collision of two spheres, the new velocities are corrected to move apart faster than the growth rate $d\sigma/dt$, whereupon

$$\underline{\mathbf{v}}_i \text{ (new)} = ( \underline{\mathbf{v}}_i\text{(old)} \underline{\pm} \underline{\mathbf{v}}_{ii}\underline{\mathbf{r}}_{ij} ) (1+ d\sigma/dt) \qquad \text{equation (5)}$$

Equation (1) is integrated numerically to determine $V_\infty$ using a finite time difference predictor algorithm given a starting volume $V_i$ of an equilibrated fluid configuration. At each $\Delta t$, the unknown $V_\infty$ is predicted at the beginning, and then corrected at the end, from the mean pressure. The work done on the system is continually removed by velocity rescaling to maintain total kinetic energy at $3Nk_BT/2$. As close-packing is approached the volume $V_\infty$ converges, as the pressure diverges.

Figure 2 shows NEMD results for the close-packed state over eight orders-of-magnitude for the rate constant $k_1$. At very low $k_1$ the system crystallizes, perfect FCC for N=500, $V_\infty = 2^{-½} N\sigma^3$. For faster $k_1$ increasingly defective predominantly crystalline states are obtained. There is a narrow intermediate range of steeply increasing residual volume, which may be heterogeneous for large systems, followed by glasses with nucleites, and eventually an RCP plateau. RCP states can be obtained at a sufficient $k_1$, when the system can equilibrate locally, but has no time and/or space in which to crystallize, or even nucleate.

The residual volume of RCP fluctuates for small systems. As N increases, however, there is a slight number dependence, and the RCP state converges to a residual volume close to 0.120. The data is insufficient to obtain the N-dependence of RCP by this process at this stage. Amorphous, jammed states of lower RCP fraction, are seen for $k_1 > 10$. The finite–difference algorithm for $d\sigma/dt$ (equations 1-3) becomes unstable around $k_1 \sim 20$.

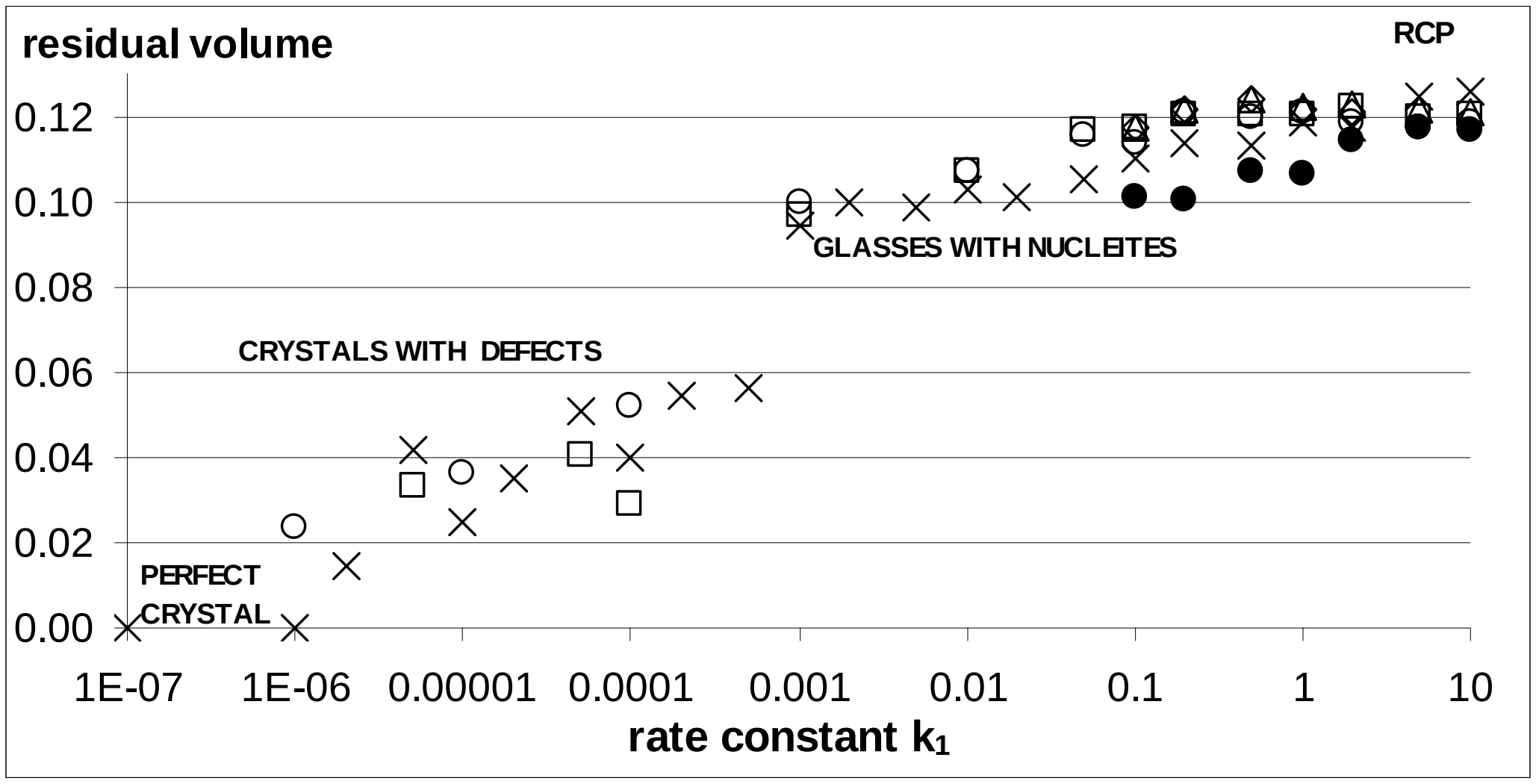

**Figure 2 Residual volumes ($\Delta V/N\sigma^3$) of close-packed states of hard spheres as a function of the compaction rate constant in dimensions of $(m\sigma^2/k_BT)^{1/2}$. In the limit of slow compaction, crystallization occurs ($V_{residual}$ = 0 for fcc crystal) Crosses are results for N=500 starting at $\rho_i\sigma^3$=0.90. All other data points correspond to N=1000 for different starting densities $\rho_\iota\sigma^3$ = 0.95 (circle); 0.75(square); 0.55 (diamond) and 0.35 (triangle); solid circles N=8000. Residual volume of 0.12 $N\sigma^3$ corresponds to a RCP fraction 0.633.**

RCP states of the hard-sphere glass obtained by the above irreversible process, can be accessed reversibly. The thermodynamic pathway to RCP makes use of a single-occupancy (SO) system. The idea of a reversible pathway, using a FCC single-occupancy (FCCSO)-cell system was originally introduced [14], to determine the hard-sphere freezing transition. The center of each sphere is confined within the walls of its cell, as defined by the primary Voronoi polyhedra. Starting with the ideal gas, the p-V equation-of-state provides a reversible path from fluid to crystal which enables thermodynamic properties such as the relative free energy, and hence the crystallization transition pressure, to be calculated.

The simple-cubic SO-system (SCSO) provides a reversible path to a homogeneous amorphous glass. Its close-packed crystal configuration ($\rho\sigma^3 = 1$) is unstable with respect to slip in all directions, and nucleation to alternative crystal structures is prohibited. The system, shows no phase transition, as it compacts smoothly and reversibly from the SO-ideal gas at low density, through its own SCSO lattice fraction of $y = \pi/6$, to a RCP structure at $y = 0.637$, for N= 8000 spheres (Figure 3). The SCSO p–V equation-of-state appears to be everywhere continuous. It becomes equal to the hard-sphere fluid in the limit of both low density, where it obeys the ideal gas law, and at high density, obeying the SCFV equation as it approaches RCP reversibly.

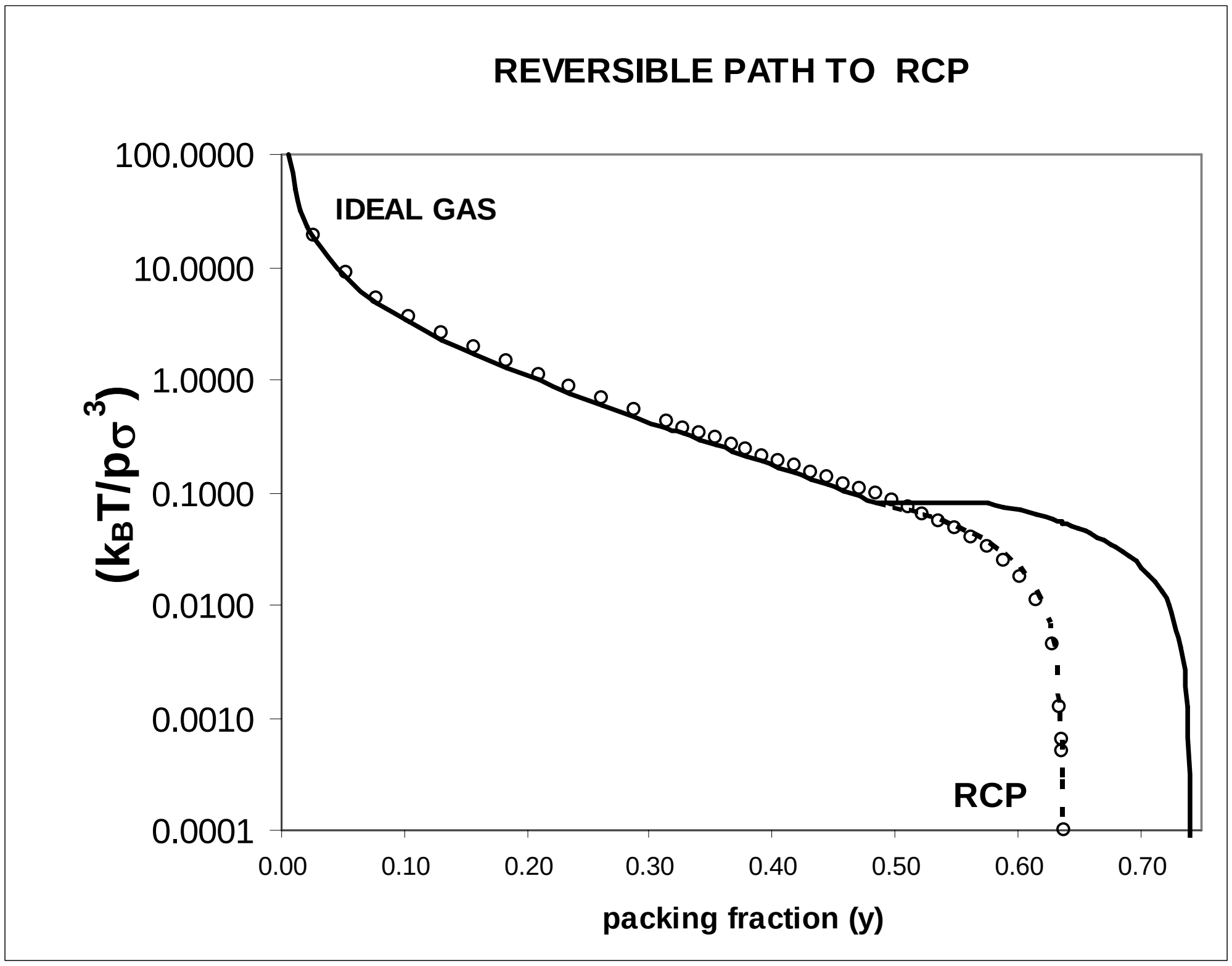

**Figure 3 From the ideal gas (y=0) to an ideal glass (y=0.637) Equations-of-state for the amorphous simple-cubic single-occupancy (SCSO) reversible path to RCP (open circle data points N=8000), The solid black line is the equilibrium hard-sphere fluid and crystal; the SCSO-system, and the rapidly quenched unconstrained metastable hard-sphere fluid (dashed line) become the same in the limit of RCP as $k_BT/p\sigma^3 \rightarrow 0$.**

Communal entropy can be defined as the entropy difference between a SO-cell model and the free fluid at the same temperature and volume ($\Delta_{SO}S_{T,V}$). In the low density ideal gas limit, all SO systems approach the ideal-gas equation and the communal entropy is $\Delta_{SO}S_{T,V}$ is exactly $Nk_B$ . This result provides a check on the reversibility of the SCSO pathway to RCP. If the two states, equilibrium SCSO and metastable HS-glass, at RCP are indeed identical, the communal entropy must disappear along the reversible pathway between y = 0 and RCP in Figure 3. The integral of the difference between SCSO and the free hard-sphere fluid and its metastable branch from ideal gas to RCP must equate to the communal entropy, i.e.

$$\Delta S / Nk_B = \int_{\infty}^{RCP} (\Delta p/T)\, dV - 1 = 0 \qquad \text{equation (6)}$$

At RCP is approached, the effect of the SO-cell walls becomes negligible, $\Delta p$ approaches zero and the communal entropy disappears at RCP.

Small finite systems exhibit values of properties which depend on the system size. For real solid spheres in rigid containers, it was found that the maximum packing fraction increases as $N^{1/3}$. In the case of computer models with periodic boundary conditions, the N-dependence is weaker. A plot of the RCP fraction against $N^{-1/3}$ (Figure 4) can be fitted to a linear trendline that interpolates at N = ∞ to 0.6333±0.0005. A near identical result to that obtained by Scott [2] almost 50 years ago with up to 20000 steel balls in a copper cylinder!

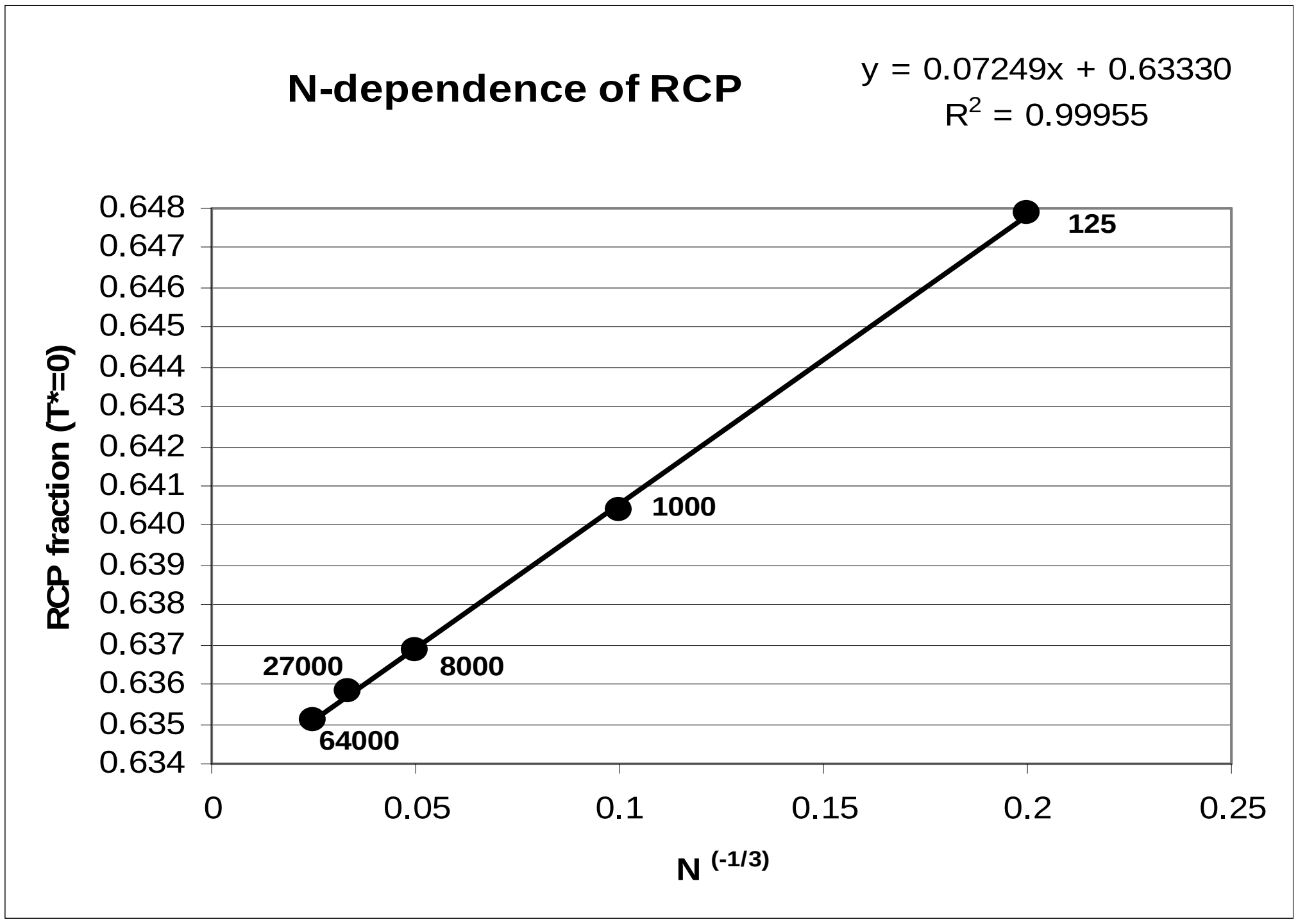

**Figure 4 RCP fraction for the simple-cubic single-occupancy (SCSO) system for N spheres from N=125 to N=64000 with periodic boundaries. A best-fit trendline of y(RCP) against $(I/N)^{1/3}$ interpolation to the infinite system limit gives y (RCP : N→∞ ) = 0.6334± 0.0005**

A knowledge of the SCSO p-V equation-of-state permits definition and determination of its thermodynamic properties. Metastable amorphous solid states are characterized by thermodynamic excess properties relative to the crystal at the same temperature and pressure. The residual volume is obtained from $V_\infty$(RCP)–$V_0$ (FCC). The RCP excess entropy can be obtained, together with some insights into its origins, from the difference between the SCSO and FCCSO p-V equations-of-state.

From the 1st and 2nd law of thermodynamics, any change in Gibbs energy is given by the differential equation dG = Vdp at constant T . Also for hard spheres the enthalpy change dH = pdV. Since, dG = dH - TdS the entropy difference between FCCSO and SCSO at the same T and p is

$$\Delta S_{T,p} = (p/T)\,\Delta V_{T,p} - \int \Delta V_{T,p}\, dp/T \qquad \text{equation (7)}$$

Not only are the two equations-of-state identical in the low density limit, but also very close, almost the same, for all pressures below ~ 8.5($\sigma^3$/kT) i.e. that of the ordering transition in the FCCSO equation-of-state. Moreover, for pressures greater than around 15 ($\sigma^3$/kT) the residual volume of the SCSO becomes constant with further increase in pressure. Beyond this pressure the two terms in equation (7) exactly cancel. This means that both systems have identical thermal expansivities and heat capacities in the high pressure region for $p\sigma^3/kT >\sim 15$ as the pressures diverge towards infinity. A value $\Delta S_{residual} = 1.08 \pm 0.02\ Nk_B$ is obtained for RCP entropy by numerical evaluation of equation (7). The residual entropy is mainly the entropy of the hard-sphere fluid-FCC melting transition. This implies a connection between the equilibrium thermodynamic states of the hard-sphere fluid at freezing, and the non-equilibrium RCP state.

In answer to the question: "what is RCP ?" [10] , it is postulated that RCP state is an analytic extrapolation of the equilibrium fluid equation at freezing $\rho_f\sigma^3$ (0.95), to densities beyond freezing, past the limit of the metastable branch (1.035), via the unstable region of spontaneous nucleation, to a state of infinite pressure (1.2155) , assuming all its p-V derivatives at $\rho_f$ are continuous. This was first suggested by Le Fevre [15] and later supported by computer experiments [7].

"Is RCP well defined?" [10,11] Yes: provided crystal nucleation is prohibited, an upper-bound RCP fraction becomes independent of a rate constant for a well-defined irreversible compaction process, where it is independent of $V_i$ and also only weakly N-dependent. The same RCP state can be accessed via the SCSO-system.

Is RCP fraction a minimum or a maximum ? [7,8] . RCP is definitely not a minimum density for homogeneous amorphous close packing. Alternative compaction processes and athermal algorithms, lead to homogeneous amorphous “jammed states” with lower densities [9,13]. The present results indicate a local maximum in density and entropy (for given N), provided nucleation is prohibited. For thermal metastable states in the vicinity of RCP, thermodynamics requires that small displacements must increase the Helmoltz free energy, i.e. $dA = -TdS = pdV \geq 0$ . RCP is a local, but not global, minimum in free energy.

Finally, “why is RCP reproducible? [12] . When RCP is measured or computed very precisely, say to an accuracy of less than 0.1%, its not so reproducible! RCP fractions from both granular and computer experiments can vary with the compaction process, the starting configuration, and the number of spheres, but all within a narrow range of around 1%. Since RCP is a local maximum entropy in amorphous phase space, systems of spheres with kinetic energies must gravitate in a direction towards it, given the opportunity. To obtain RCP however, local relaxation must be allowed, and crystal nucleation must be disallowed, in any irreversible compaction process.

**Acknowledgements**

I wish to than Professor J L Finney of University Colege London for his constructive advice, and positive cooments and advice in the presubmission of this letter.

**Literature Cited**

1. Bernal, J.D., “A Geometrical Approach to the Structure of Liquids” Nature **183,** 141-147 (1959)

2. Scott, G.D., Nature, “Packing of Equal Spheres”, **188** 910-911 (1960)

3. Bernal J D and Mason J., “Coordination of Randomly Packed Spheres”, Nature **188** 912-913 (1960)

4. Finney, J.L., “Random packings and the structure of simple liquids I The geometry of random close packing” Proc. Roy. Soc. (London) A319 479-493 (1970)

5. Gotoh K., and Finney J.L., “Statistical Geometrical Approach to Random Packing Density of Equal Spheres”, Nature **232** 202-205 (1974)

6. Finney J. L., “Modeling the Structures of Amorphous Metals and Alloys”, Nature, 266 309 (1977)

7. Woodcock, L. V.,” Glass Transition in the Hard-Sphere Model and Kauzmann’s Paradox”, Ann. NY Acad. Sci. **371** 274-298 (1981)

8. Berryman, ”Random Close Packing of Hard Spheres and Disks”,, J.G., Phys Rev **A27** 1053-1061 (1983)

9. Jodrey, W.S. and Tory, E.M., ”Computer Simulation of Close Random Packing of Equal Spheres”, Phys Rev. **A 32** 2347 (1985)

10. Molecular Physics Correspondent, “What is Random Packing?” Nature **239** 488-489 (1972)

11. Torquato, S. , Truskett, T. M. and Debenedetti, P.G. ,” Is Random Close Packing of Spheres Well-Defined? Phys. Rev. Lett. **84** 2064-2067 (2000)

12. Kamien, R.D. and Liu, A. J., “Why is Random Close Packing Reproducible?” Phys. Rev. Lett. **99** 15501 (4) , (2007)

13. Auer, S. A. and Frenkel, D., “Prediction of Absolute Crystal Nucleation Rates in Hard-Sphere Colloids” Nature **409** 1020-1023 (2001)

14. Hoover W.G. and Ree, F.H. ”Communal Entropy and Freezing Transition of the Hard-sphere Fluid”, J. Chem. Phys. **49** 3609-3619 (1968)

15. Le Fevre, E. J., “Equation of State for Hard –Sphere Fluid”, Nature, Phys. Sci., **235** 20 (1972)